\documentclass{article}

\usepackage{cite} 
\usepackage{url}  
\usepackage{ifthen}  
\usepackage{multicol}   
\usepackage[utf8]{inputenc} 
\newboolean{publ}

\usepackage{amsmath}
\usepackage{tikz}
\usetikzlibrary{shapes,arrows}
\tikzstyle{S}=[draw, fill=blue!20, minimum size = 2.47em, text
width=2cm, text centered]
\tikzstyle{phi}=[draw, fill=blue!10, minimum size = 2.47em, text
width=2cm, text centered]
\tikzstyle{bigS}=[draw, fill=blue!20, minimum width=23em, text
width = 8.5cm, minimum height=5em]
\tikzstyle{I}=[draw, fill=red!20, minimum size = 2.47em, text
width=2cm, text centered]
\tikzstyle{R}=[draw, fill=yellow!20, minimum size = 2.47em, text
width=2cm, text centered]
\tikzstyle{empty}=[minimum size=0em]

\newcommand\T{\rule{0pt}{2.6ex}}  
\newcommand\B{\rule[-1.2ex]{0pt}{0pt}}
\newcommand\Ro[1][\relax]{\ifx\relax#1 \ensuremath{\mathcal{R}_0}
  \else \ensuremath{\mathcal{R}_{0,#1}} \fi}
\newcommand{\order}{\mathcal{O}}
\newcommand{\ave}[1]{\left \langle #1 \right \rangle}

\usepackage{hyperref}

\title{Epidemics on networks with large initial conditions or changing
structure}
\author{Joel C. Miller}

\begin{document}



\maketitle


\begin{abstract}
  \emph{Background:} Recently developed techniques to
  study the spread of infectious diseases through networks make
  assumptions that the initial proportion infected is
  infinitesimal and the population behavior is static throughout
  the epidemic.  The models do not apply if the initial
  proportion is large (and fail whenever $\Ro<1$), and cannot
  measure the impact of an intervention.  
    \emph{Methods:}  In this paper we adapt ``edge-based
  compartmental models'' to situations having finite-sized
  initial conditions.  
    \emph{Results:} The resulting models remain simple and
  accurately capture the effect of the initial conditions.  It
  is possible to generalize the model to networks whose
  partnerships change in time.
    \emph{Conclusions:}  The resulting models can be applied
  to a range of important contexts.  The models can be used to
  choose between different interventions that affect the disease
  or the population structure.
\end{abstract}

\ifthenelse{\boolean{publ}}{\begin{multicols}{2}}{}

\section*{Background}
The mathematical study of infectious disease spread has contributed significantly to our ability to design effective interventions to reduce disease spread.  Most of the earliest models were based on the assumption that disease transmission occurs as a Poisson process and each transmission reaches an individual chosen randomly from the population.  This implicitly assumes that partnership duration is very brief.  These models have been modified to account for a number of different effects, such as demographic groups~\cite{andersonmay}.

More recently, attempts have been made to incorporate the ``network''
structure of the population (see,
\emph{e.g.},~\cite{bansal:individual}).  Typically these focus on
trying to understand the role played by ``high-degree'' individuals
(those individuals with many contacts).  Typically these studies come
in one of two flavors: they either continue the assumption of fleeting
partnerships (the disease spreads slowly compared to partnership
turnover)~\cite{andersonmay,may:hivdynamics, may:dynamics,moreno,pastor-satorras:scale-free}, or they take the opposite limit in which the
partnership network is static (the disease spreads quickly compared to
partnership
turnover)~\cite{volz:cts_time,decreusefond:volz_limit,house:insights,newman:spread,meyers:sars,meyers:directed,lindquist}

Recent work has shown that for susceptible-infectious-recovered (SIR)
models, it is possible to unify these two approaches with an
``edge-based compartmental model'' that
allows partnership duration to range continuously from zero to
infinite~\cite{miller:ebcm_overview} [for
susceptible-infectious-susceptible (SIS) models, the picture is more
complicated, see for example~\cite{kiss:dynamic}].  The resulting models are
low-dimensional and contain many standard models as special
cases~\cite{miller:ebcm_hierarchy}.  Unfortunately, these models are
derived under the assumption that the initial proportion infected is
infinitesimally small (while the absolute number infected is
sufficiently large that the dynamics are deterministic).  It is
assumed that by the time the equations are used, any early transients
have died away.  A consequence of this assumption is that the models
break down if $\Ro<1$ or if the initial proportion infected is not
negligible.

The failure if the initial proportion infected is not negligible was
observed by~\cite{lindquist}.  This paper used an early (static
network) version of the equations of~\cite{miller:ebcm_overview}
from~\cite{volz:cts_time} and compared them with simulation.  A small
discrepancy in final sizes was noted.  This discrepancy was not
present for equations of~\cite{ball:network_eqns} a system requiring
$\order(M)$ equations where $M$ is the maximum degree or for another
system presented in~\cite{lindquist} which required $\order(M^2)$
equations.  

In this paper we will see that the discrepancy found
by~\cite{lindquist} results from the fact that the initial proportion
infected was nonzero.  We first show how to correct the equations to
accommodate a non-negligible initial proportion infected.  We then
compare the model with simulations using different assumptions about
how the initial infected individuals are chosen.  Following that, we
look at how the equations can be used to account for a change in
population structure or disease parameters during the epidemic, such
as might occur if an intervention is implemented.  We then analyze the
dynamical structure of the equilibria found in the equations, showing
how the epidemic threshold is modified when a non-negligible
proportion of the population is infected.  Finally we discuss the
assumptions underlying our approach and consequences when those
assumptions fail.

\section*{Methods}
We modify the approach of~\cite{miller:ebcm_overview} which assumed an
infinitesimal initial proportion infected.  We adapt the approach to
consider a wide range of possible initial conditions.  We assume that
the dynamics of the epidemic may be treated as deterministic, which
means we assume the population is very large and the initial number
infected is large enough for the epidemic to behave
deterministically.  If stochastic effects are still important but
$\Ro>1$, then these equations may become accurate at a later time once
sufficient numbers are infected.

We assume the population consists of $N \gg 1$ individuals.  Each is
assigned a degree $k$ independently of others, with probability $P(k)$
where $P$ defines a probability distribution on the (non-negative)
integers.  The network is wired together using the ``Configuration
Model'' (or ``Molloy-Reed'') approach~\cite{newman:spread,MolloyReed}:
each individual is assigned a number of stubs (or half-edges) equal to
its degree.  Pairs of stubs are then wired together to form
edges/partnerships.  It is likely that this algorithm produces a
handful of self-loops or repeated edges, but generally the frequency
of these goes to zero like $1/N$.

We define a \emph{test individual} $u$ to be a randomly chosen
individual.  Because we assume that the spread is deterministic, this
means that the probability $u$ is in a given state is equal to the
proportion of the population that is in that state.  So we focus on
calculating the probability $u$ is susceptible, infected, or
recovered.  We modify $u$ so that it does not transmit to any of its
partners if ever infected.  This assumption does not affect the
probability $u$ is in any given state, but it does prevent a
correlation between the status of different partners which would be
caused by infection traveling through $u$.  This allows us to treat
the partners of $u$ as independent.  It is important to note that this
assumption has no impact on the probability $u$ is in any given state
and therefore, it does not affect our calculation of the proportion of
the population in each state.  Further discussion of the test
individual is in~\cite{miller:final}.

\subsection*{Variables}
We introduce our variables in table~\ref{table:variables}.  The
starting point is the test individual.  The remaining
variables can be broadly divided into three groups.  $S$, $I$, and $R$
denote the proportion of the population in each state, or equivalently the
probability that the test individual $u$ is in each state.  $\theta$,
$\phi_S$, $\phi_I$, and $\phi_R$ give information about the
probability a partner of $u$ has a given status and the probability
the partner has transmitted to $u$.  $P$ gives information
about the possible degrees of $u$ or its partners, while $S(k,0)$
gives information about the probability $u$ is initially susceptible.
Given $P$, $\theta$ and $S(k,0)$, we define $\psi(x) = \sum S(k,0)P(k)
x^k$, so that $\psi(\theta(t))$ is the probability $u$ is
susceptible.  By noting that $\psi(\theta(t))=S(t)$, we will be able
to close our system of equations.

\begin{table}
\begin{center}
\begin{tabular}{c|c}
  Variable & Definition\\
  \hline\hline\hline
  Test Individual $u$ & \parbox{0.6\textwidth}{\T{}A randomly chosen member of the population who
    is prevented from causing infection.\B{}}\\\hline\hline\hline
  $S(t)$ & \parbox{0.6\textwidth}{\T{}The proportion of the entire population that is susceptible.\B{}}\\\hline
  $I(t)$ & \parbox{0.6\textwidth}{\T{}The proportion of the entire population that is infected.\B{}}\\\hline
  $R(t)$ & \parbox{0.6\textwidth}{\T{}The proportion of the entire
    population that is recovered.\B{}}\\\hline\hline\hline
  $\theta(t)$ & \parbox{0.6\textwidth}{\T{}The probability a random
    partner $v$ of $u$ which did not transmit to $u$ by $t=0$ has
    not transmitted to $u$ by time $t$.\B{}}\\\hline
  $\phi_S(t)$ & \parbox{0.6\textwidth}{\T{}The probability a random partner $v$ which did not transmit to $u$ by $t=0$ is
    susceptible at time $t$.\B{}}\\\hline
  $\phi_I(t)$ & \parbox{0.6\textwidth}{\T{}The probability a random partner $v$ which did not transmit to $u$ by $t=0$ is
    infected at time $t$ but has not transmitted to $u$.\B{}}\\\hline
  $\phi_R(t)$ & \parbox{0.6\textwidth}{\T{}The probability a random partner $v$ which did not transmit to $u$ by $t=0$ is
    recovered at time $t$ and never transmitted to $u$.\B{}}\\\hline\hline\hline
  $P(k)$ & \parbox{0.6\textwidth}{\T{}The probability an individual has
    degree $k$.\B{}}\\\hline
  $\ave{K}= \sum_k kP(k)$ & \parbox{0.6\textwidth}{\T{}The average degree.\B{}} \\ \hline
  $S(k,0)$ & \parbox{0.6\textwidth}{\T{}The probability an individual
    with degree $k$ is initially susceptible.\B{}}\\\hline
  $\psi(\theta(t)) = \sum_k S(k,0)P(k)\theta(t)^k$
  & \parbox{0.6\textwidth}{\T{}The probability that the test individual
    $u$ is susceptible at time $t$.  In a large population this should equal $S(t)$.\B{}}\\
  \hline\hline\hline
\end{tabular}
 \end{center}
\caption{\small The variables we need to calculate the epidemic dynamics.  In
  all of these $u$ is a test individual: randomly chosen from the
  population and modified so that it cannot infect others,
  although it can become infected.}
\label{table:variables}
\end{table}

The main distinction between this approach and the previous
approach~\cite{miller:ebcm_overview} is that we use just the initially
susceptible individuals to define $\psi$ while the earlier work
assumed $S(k,0)=1$ for all $k$.  In this case, $\psi$ is a probability
generating function for the degree distribution.

\subsection*{Equation Derivation}
We will find a closed system of equations based on these variables.
We begin by looking at $S(t)$.  If the test individual $u$ has degree
$k$ and is susceptible at $t=0$, then the probability it is
susceptible at some later time is $\theta(t)^k$.  If we do not know
$k$ or whether $u$ is susceptible at $t=0$, then the probability $u$
is susceptible at time $t$ is sum over all $k$ of the product of the
probability $u$ is initially susceptible $S(k,0)$ with the probability
$u$ is still susceptible $\theta^k$.  We have $S(t)=\sum_k S(k,0)P(k)
\theta(t)^k = \psi(\theta(t))$.  Thus we conclude
\[
S(t) = \psi(\theta(t))
\]
Given our initial conditions on $I$ and $R$, we know that $R$ solves
$\dot{R} = \gamma I$.  We also have a conservation rule that
$S+I+R=1$, so $I=1-S-R$.  Thus our equations are
\begin{align*}
S &= \psi(\theta)\\
I &= 1-S-R\\
\dot{R} &= \gamma I
\end{align*}
Assuming $\theta(t)$ is known, then this system completely defines
$S$, $I$, and $R$.  This is shown in the flow diagram in
figure~\ref{fig:indflow}. 

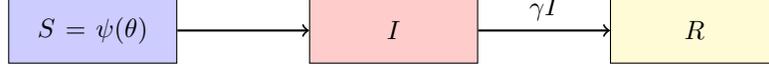
\begin{figure}
 \begin{center}
 \begin{tikzpicture}
 \node [S] at (0,0) (S) {$S=\psi(\theta)$};
 \node [I] at (4,0) (I) {$I$};
 \node [R] at (8,0) (R) {$R$};
 \path[->, thick, left] (S) edge node {}(I); 
 \path[->, thick, above] (I) edge node {{$\gamma I$}} (R);
 \end{tikzpicture}
 \end{center}
\caption{\small Flow diagram showing the flux of individuals between the
  different compartments.  Because we have an explicit expression for
  $S$, if we know $\theta$ we do not need to explicitly determine the
  flux from $S$ to $I$.}
\label{fig:indflow}
\end{figure}

In order to close this system of equations we need an equation giving
$\theta$.  Recall that $\theta(t)$ is the probability a partner $v$
of $u$ which had not yet transmitted to $u$ by time $0$ has not
transmitted by time $t$.  This can be broken in three disjoint
sub-compartments.  If
$v$ has not yet transmitted to $u$ by time $t$ it is either
susceptible, infected, or recovered.  Thus, setting $\phi_S$ to be the
probability that $v$ is still susceptible, $\phi_I$ to be the
probability $v$ is infected but has not transmitted to $u$, and
$\phi_R$ to be the probability $v$ has recovered and did not transmit
to $u$, we conclude that $\theta = \phi_S+\phi_I+\phi_R$.  It is
straightforward to see that $\dot{\theta} = -\beta \phi_I$.  So if we
can find $\phi_I$ in terms of $\theta$, then we arrive at a single
equation for $\theta$, which can be used to provide $\theta$ for the
$S$, $I$, and $R$ equations.

To do this, we use the fact that $\phi_I = \theta- \phi_S -\phi_R$ and
find $\phi_S$ and $\phi_R$ in terms of $\theta$.  We turn to
figure~\ref{fig:edgeflow}.  The recovery rate is $\gamma$
and the transmission rate is $\beta$, so we have $\dot{\phi}_R =
-\gamma \dot{\theta}/\beta$.  We can integrate this, and using the
fact that $\theta(0)=1$ we find $\phi_R = \gamma(1-\theta)/\beta +
\phi_R(0)$.  To find $\phi_S$ in terms of $\theta$, we note that the
probability $u$ has an edge to a node which is susceptible at time
$t=0$ is $\phi_S(0)$.  The probability the susceptible partner has
degree $k$ is $k P(k) S(k,0)/\sum k P(k) S(k,0)$, so the probability
an initially susceptible partner is susceptible at some later time is
$\sum_k k P(k) S(k,0)\theta^{k-1} / \sum_k k P(k) S(k,0) =
\psi'(\theta)/\psi'(1)$.  Thus $\phi_S(t) = \phi_S(0) \psi'(\theta(t))/\psi'(1)$.
We arrive at
\[
\phi_I = \theta - \phi_S(0) \frac{\psi'(\theta)}{\psi'(1)} -
\frac{\gamma}{\beta}(1-\theta) - \phi_R(0)
\]
and $\dot{\theta} = -\beta \phi_I$ becomes
\[
\dot{\theta} = -\beta \theta + \beta \phi_S(0)
\frac{\psi'(\theta)}{\psi'(1)} + \gamma(1-\theta) + \beta \phi_R(0)
\]
with $\theta(0)=1$.  This completes our system.

\begin{figure}
 \begin{center}
 \begin{tikzpicture}
 \node [bigS] at (0,0) (Btheta) {$\hspace{8.25cm}\theta$\\[1cm]~};
 \node [phi] at (0,0) (phiI) {$\phi_I$};
 \node [phi] at (-3,0) (phiS) {$\phi_S=\phi_S(0)\frac{\psi'(\theta)}{\psi'(1)}$};
 \node [phi] at (3,0) (phiR){$\phi_R$};
 \node [I] at (0,-3) (OmT) {$1-\theta$};
 \path[->, thick,left] (phiI) edge node {{$\beta\phi_I$}} (OmT);
  \path[->, thick] (phiS) edge node {{}} (phiI);
  \path[->, thick,above] (phiI) edge node {{$\gamma\phi_I$}} (phiR);
 \end{tikzpicture}
 \end{center}
\caption{\small Flow diagram for the flux of partners through different
  states.   The top three boxes $\phi_S$, $\phi_I$, and $\phi_R$
  represent the different states the partner can be in if it has not
  transmitted.  The lower box $1-\theta$ is the probability the
  partner has transmitted.}
 \label{fig:edgeflow}
\end{figure}
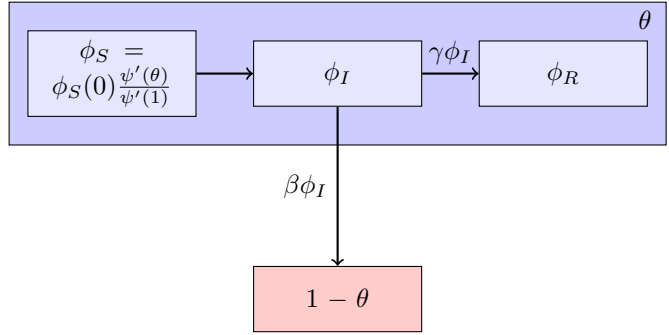

Our final closed system of equations is
\begin{align}
\dot{\theta} &= -\beta \theta + \beta \phi_S(0) \frac{\psi'(\theta)}{\psi'(1)} + \gamma(1-\theta) + \beta \phi_R(0)\label{eqn:sys1}\\
\dot{R} &= \gamma I \, , \qquad S=\psi(\theta) \, , \qquad I=1-S-R \label{eqn:sys2}
\end{align}
where $\theta(0) = 1$, and $R(0)$ is given by the initial conditions.
These equations lead to earlier equations
of~\cite{miller:ebcm_overview,volz:cts_time,miller:volz} if $\Ro>1$
and $1-\phi_S(0)$, $\phi_I(0)$, $\phi_R(0)$ and
$1-\theta(0)$ are all infinitesimally small.

\subsubsection*{Generalizing the model}
In~\cite{miller:ebcm_overview}, a number of generalizations to various
dynamic network structures were considered.  The basic approach we
have used here can be applied to any of those generalizations.  

\subsubsection*{Final size relation}
The final size relation assuming small initial condition is
well-known~\cite{newman:spread}.  The final size relation for larger
initial conditions has recently been found~\cite{miller:final} in a
more general case not assuming constant transmission and recovery
rates.  It can be derived easily for this model by setting
$\dot{\theta}=0$.
We find
\begin{align}
\theta(\infty) &= \frac{\beta\left[\phi_S(0)\frac{\psi'(\theta(\infty))}{\psi'(1)} + \phi_R(0)\right] +
\gamma}{\beta+\gamma}\label{eqn:fs1}\\
R(\infty) &= 1-\psi(\theta(\infty)) \label{eqn:fs2}
\end{align}

\section*{Results and Discussion}
\subsection*{Model Validation}
In this section we compare our model with simulations for populations
which satisfy the Configuration Model/Molloy-Reed model assumptions.
Although an earlier version of our equations was found to have minor
discrepancies~\cite{lindquist}, we show that once we appropriately
account for the initial condition, the calculation becomes correct.

\subsubsection*{Final Size Comparison}
To show that our new equations accurately calculate the
impact of the initial conditions, we first consider epidemic spread in
networks with the same degree distribution as in~\cite{lindquist}, but
with varying numbers infected and varying population sizes.  We then
consider the impact of selecting high or low degree nodes as the
earliest infected individuals, using networks whose degree
distributions more clearly show the impact of biased selection of the
initial individuals.

We run a large number of simulations for each number of initial
infections.  For each simulation we generate a new network.  Our
simulation technique is similar to those recently described
by~\cite{noel:eh,ball:network_eqns,decreusefond:volz_limit}.  In the Configuration Model
framework, each node is assigned a degree, nodes are given stubs (or
half-edges), and then stubs are randomly paired together.  In the
simulations we use, each node is assigned a degree, nodes are given
stubs, and then the disease begins to spread in the network before
stubs are paired.  Each time the disease transmits along a stub that
stub is randomly paired with another as yet unpaired stub.  If the
partner is susceptible, then it becomes infected.  If not, nothing
happens.  Once stubs are paired they remain in their edge.  This
approach is equivalent to constructing the network in advance
and then following the disease, but it is more efficient
because it only constructs those parts of the network the disease
traces.

\paragraph{Randomly selected initial infections}
We first consider varying numbers of randomly chosen infected
individuals.  In figure~\ref{fig:lindquistcomp} we take the degree
distribution from~\cite{lindquist}.  We have $P(1) = 18.118\times
10^{-3}$, \ $P(2) = 72.536\times 10^{-3}$, \ $P(3) = 145.222\times
10^{-3}$, \ $P(4) = 194.589\times 10^{-3}$, \ $P(5) = 195.962\times
10^{-3}$, \ $P(6) = 156.857\times 10^{-3}$, \ $P(7) = 105.280\times
10^{-3}$, \ $P(8) = 59.713\times 10^{-3}$, \ $P(9) = 30.066\times
10^{-3}$, and $P(10) = 21.657\times 10^{-3}$.

\begin{figure}
\includegraphics{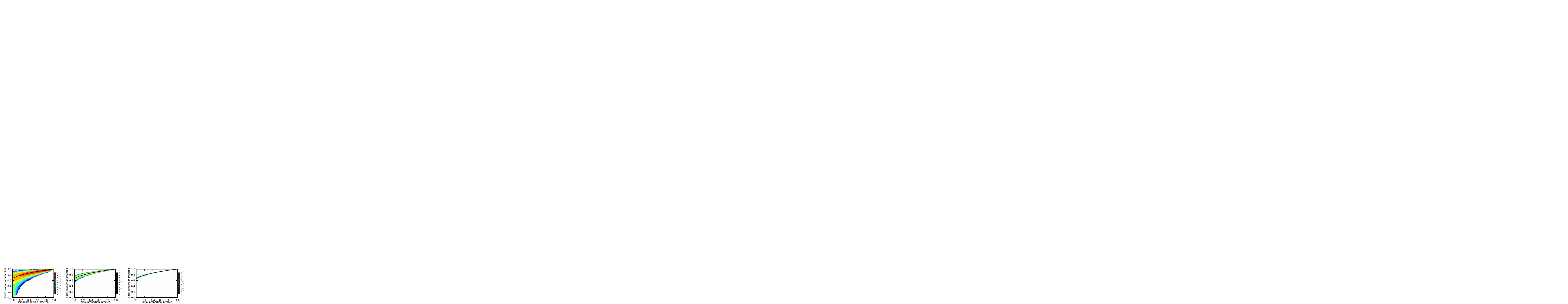}
\caption{\small Results of simulations for $100$, $1000$, and $10000$
  individuals.  The solid curve gives our prediction for the final
  sizes of epidemic in a large population.  Colors are log scale
  giving frequency of that particular epidemic size.  In the case of
  $100$ individuals we take every possible initial condition from $1$
  to $100$ infections and perform $50000$ simulations.  In the case of
  $1000$ individuals, we take every possible initial condition from
  $1$ to $1000$ infections and perform $11500$ simulations.  In the
  case of $10000$ individuals, however, we only look at multiples of
  $10$ initial infections, performing $3000$ simulations at each
  level.  To show that the number of simulations performed accurately
  capture the full range, we show an increased number of simulations
  from $17.5\%$ to $22.5\%$, performing 2000000, 50000, and 10000
  simulations.  It is clear this increased number of simulations has
  no significant impact.}
\label{fig:lindquistcomp}
\end{figure}

We randomly select a proportion $\rho$ of the population to 
initially infect.  We have $S(k,0)=1-\rho$ for all $k$, so $\psi(x) =
(1-\rho)\sum_k P(k) x^k$.  Similarly we have $\phi_S(0)=1-\rho$.
Because the epidemic begins with no recovered individuals, we take
$\phi_R(0)=0$.   We take $\beta = 0.1$ and $\gamma = 0.2$

We take populations of $100$, $1000$, and $10000$ and perform many
simulations.  To compare with our predictions, we consider the final
sizes observed, using the final size relation of
equations~\eqref{eqn:fs1} and~\eqref{eqn:fs2} to compare with
simulations. The equations are derived in the infinite population
limit, but we see that even with populations of only $100$ they give a
good prediction of the observed behavior.  As the population size
increases, the noise becomes less significant and the simulations
collapse tighter around the prediction.

\paragraph{Biased initial infections}
\begin{figure}
\includegraphics{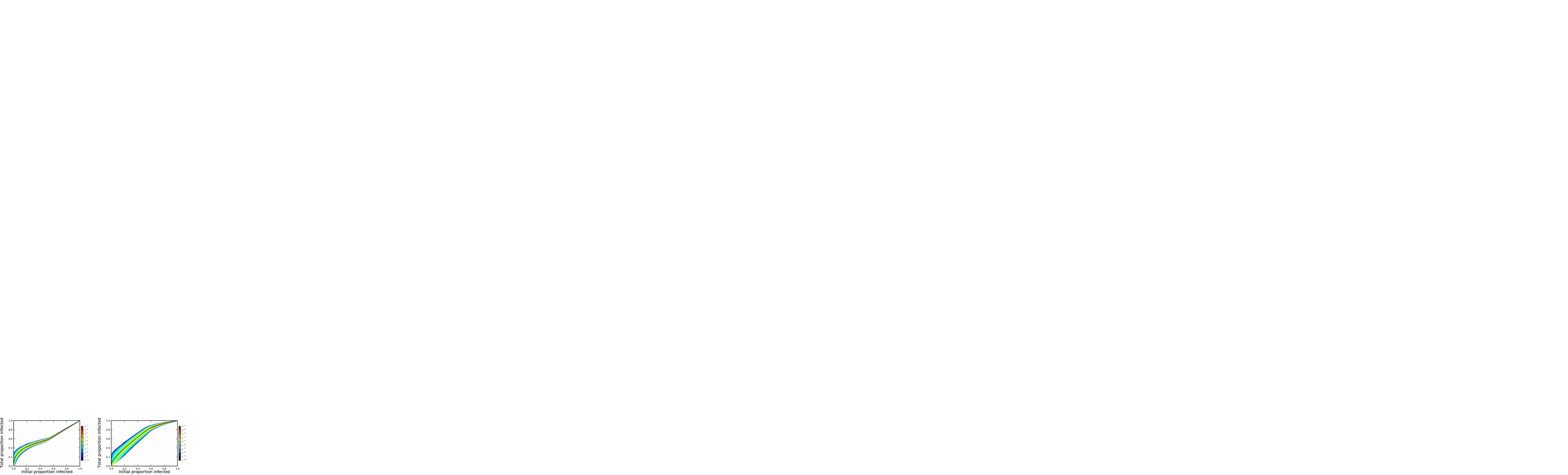}
\caption{\small Epidemic final sizes in population of $1000$ individuals with
 half having degree $9$ and half with degree $1$.  The disease
 parameters are $\beta = 0.1$, \ $\gamma = 0.6$.  Results of
 simulations having initial infections chosen with probability
 proportional to square of degree (left) or inverse square of degree
 (right).  For each initial number of infections, $22500$ simulations
 were performed.  For the range between $17.5\%$ and $22.5\%$,
 $200000$ simulations were performed to give insight into how well
 resolved the distribution is.}
\label{fig:biased_IC}
\end{figure}

 To show that the approach we have derived can also be applied to cases
where the initial infected individuals are selectively chosen based on
their degree, we use a different degree distribution which helps
highlight the effect.  We take $P(1)=P(9)=1/2$.  We consider two
options.  In the first approach, individuals with higher degree are
preferentially selected.  To do the selection, we choose an individual
with probability proportional to the square if its degree, and infect
it.  We repeat this, until a proportion $\rho$ of the population
is infected.  In the second approach individuals are chosen with probability
proportional to the square of their inverse degree until a proportion
$\rho$ is infected.  We take $\beta=0.1$ and $\gamma = 0.6$.

Using these rules, we clearly see that $S(k,0)$ is not uniform.
Instead, for the case where individuals are selected with probability
proportional to their squared degree, we find that $S(k,0)=
\alpha^{k^2}$ where $\alpha$ solves $\sum_k P(k) \alpha^{k^2} =
1-\rho$.  We find $\phi_S(0) = \sum k S(k,0) P(k) \alpha^{k^2}/\sum_k
kP(k)$.  In the case where individuals are selected with probability
inversely proportional to their squared degree, we find that
$S(k,0)= \alpha^{1/k^2}$ where $\alpha$ solves $\sum_k P(k)
\alpha^{1/k^2} = 1-\rho$, and $\phi_S(0) = \sum k S(k,0) P(k)
\alpha^{1/k^2}/\sum_k kP(k)$.

We compare predictions and simulations in populations of $1000$
individuals in figure~\ref{fig:biased_IC}.  In the limit of a
negligible initial proportion infected, the final size of epidemics in
these networks is about $4\%$.  As we increase the number of initially
infected individuals, we increase the final size because of these
individuals and because of the additional infections they lead to.  At
small amounts, increasing the number of high degree nodes has a much
larger impact on the final size because they cause more additional
infections.  However, as the amount of infection initially present is
increased this effect becomes less important: the high degree
individuals would become infected anyway.  So the largest gain in
final size comes from infecting low degree individuals who would not
receive an infection from their partners.  The ``kinks'' that occur
just above $50\%$ initially infected are because effectively all
individuals of high (left) or low (right) degree are initially
infected.

\subsubsection*{Dynamic Calculation}
We now look at the performance of the dynamic equations.  The dynamic
prediction is more easily affected by noise than the final size
prediction, so we use larger population sizes.  We again take the
degree distribution of~\cite{lindquist}.  We begin with $5\%$
infected, either randomly chosen, or chosen as before proportional to
the square of the degree.  A comparison of simulation with
calculations is in figure~\ref{fig:dynamic}.  The theory accurately
predicts the dynamics of epidemics.

\begin{figure}
\includegraphics{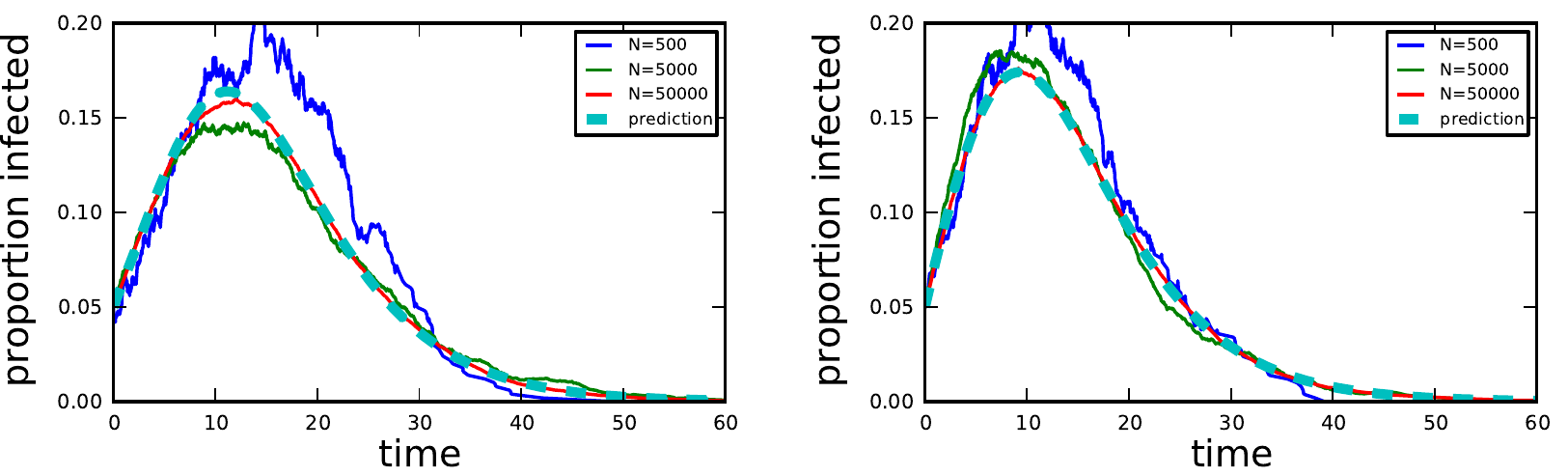}
\caption{\small A comparison of the observed and predicted number of
  infections from simulations.  Left: 5\% initially infected, chosen
  randomly  from the population.  Right: 5\% initially infected,
  chosen with probability proportional to squared degree. }
\label{fig:dynamic}
\end{figure}

\subsection*{Intervention Impact}
We can use our equations to compare the impact of several interventions.
We consider an epidemic spreading in the population, and at some
intermediate time we introduce a change in the disease or population.
Because the system changes at a time with a non-negligible amount of
infection in the population, the equations derived assuming a
negligible proportion infected fail.

Consider a population in which $P(4)=P(5)=P(6)=1/3$.  Assume we
initially infect a small, randomly chosen proportion of the
population, $\rho$ at $t=0$.  Thus we have $\psi(x) = (1-\rho)(x^4+x^5+x^6)/3$, \
$\phi_S(0)=1-\rho$, \ $\phi_I(0)=\rho$, and $\phi_R(0)=0$.  We take
$\beta=1$ and $\gamma = 1/2$.  

We consider three interventions which may be introduced at time $t_1$.
All are aimed at ``halving'' the transmission rate, but they do this
in different ways.  In mass-action based models, these would all have
the same effect.  We can clearly identify differences using our
approach.

\begin{enumerate}
\item An intervention that reduces $\beta$ by a factor of $2$.
\item An intervention that reduces $\beta$ so that per-contact transmission
  probability $\beta/(\beta+\gamma)$ is reduced by a factor of $2$.
\item An intervention that eliminates half of the partnerships
  randomly.
\end{enumerate}

The distinction between the first two comes from the fact that
partnerships have duration.  The expected number of transmissions an
individual sends to a partner is $\beta/\gamma$, but only the first is
successful.  If we use mass action assumptions however, each
transmission is to a replacement partner, and so halving $\beta$
halves the total number of transmissions. When we account for infinite
partnership duration, the expected number of transmissions remains the
same, but some partnerships transmit more than average, so others must
transmit fewer.  The probability of transmitting at least once is
$\beta/(\beta+\gamma)$.  So to reduce infection probability by a
factor requires a larger reduction to $\beta$.  Note that the work
of~\cite{miller:bounds} suggests that in Configuration Model networks
the final size of our second and third intervention will be the same,
(but that in clustered networks it will be different).

We will demonstrate our approach in all three cases, restarting the
calculations when the intervention is put into place.  In all cases,
this allows us to use the conditions at $t_1$ to predict the final
size.  We take $\psi_0(x)= \psi(x)$, \ $\theta_0$, \ $\phi_{S,0}$, \ $\phi_{I,0}$, and
$\phi_{R,0}$ to correspond to time less than $t_1$.  We use a
subscript of $1$ for times after $t_1$.  We solve the original
equations, and then use the results to initialize the second set of
variables.  

\paragraph{Case 1}
We begin by reducing $\beta$ by a factor of $2$ at time $t_1$.  Until
time $t_1$, we are solving the original equations.  By solving the
original system until $t_1$ we have $\theta_0(t_1)$.  The probability
an individual of degree $k$ is susceptible at time $t_1$ is
$S(k,t_1)=S(k,0)\theta_0(t_1)^k$.  So our new $\psi(x)$ is $\psi_1(x) = \sum_k P(k)S(k,0)
\theta_0(t_1)^kx^k = \psi(\theta_0(t_1)x)$.  We take our new
$\theta_1$ to have $\theta_1(t_1)=1$.  The intervention we are doing
has no impact on the probability a partner is in any given state.
$\phi_S$, $\phi_I$, and $\phi_R$ keep the same proportion, but are
scaled up to sum to $1$ so each is scaled by $\theta_0(t_1)$.  For
example, $\phi_{S,1}(t_1) = \phi_{S,0}(t_1)/\theta_0(t_1) =
\phi_{S,0}(0) \psi_0'(\theta_0(t_1))/\psi_0'(1)\theta_0(t_1)$.

We restart the solutions with these new values.  

\paragraph{Case 2}
The total probability of transmitting to a partner is
$\beta/(\beta+\gamma )$.  For this intervention we change $\beta$ so
that $\beta/(\beta+\gamma)$ is reduced by a factor of $2$ at time
$t_1$.  This proceeds exactly as above except that the new value of
$\beta$ must be smaller.

\paragraph{Case 3}
When we delete half the edges at random, we do not affect the
probability that a random partner is in any given state.  So the
$\phi$ variables rescale in the same way as for changing $\beta$ in
the previous cases.  However, $\psi$ undergoes a more significant
change.  As a starting point, consider $P_1(k_1)$, the probability an
individual has degree $k_1$ after edges are deleted.  This depends on
$P_0(k_0)$, the probability of having $k_0$ edges prior to deletion.
The relation is
\[
P_1(k_1) = \sum_{k_0} P_0(k_0) \binom{k_0}{k_1}  \left( \frac{1}{2}\right)^{k_0}
\]
The probability the individual has degree $k_1$ and is
susceptible is 
\[
Q(k_1) = \sum_{k_0} P_0(k_0)  S(k_0,0)\binom{k_0}{k_1}  \left( \frac{\theta_0(t_1)}{2}\right)^{k_0}
\]
So if we restart the calculations at $t=t_1$ we have $S(k_1,t_1) = Q(k_1)/P_1(k_1)$ and 
\begin{align*}
\psi_1(x) &= \sum_{k_1} S(k_1,t_1)P_1(k_1) x^{k_1} \\
&= \sum_{k_1} Q(k_1) x^{k_1}\\
&= \sum_{k_1} \sum_{k_0}
P_0(k_0)  S(k_0,0)\binom{k_0}{k_1}  \left( \frac{\theta_0(t_1)}{2}\right)^{k_0} x^{k_1}\\
&= \sum_{k_0} P_0(k_0) S(k_0,0) \theta_0(t_1)^{k_0} \sum_{k_1}
\binom{k_0}{k_1}  \left(\frac{1}{2}\right)^{k_0-k_1}\left(\frac{x}{2}\right)^{k_1}\\
&=\sum P_0(k_0)  S(k_0,0)\theta_0(t_1)^{k_0}
\left(\frac{x+1}{2}\right)^{k_0}\\
&= \psi_0\left(\theta_0(t_1) \left[\frac{1+x}{2}\right] \right)
\end{align*}
(in general if we delete edges with probability $p$ and keep with
probability $q=1-p$, then the new function is
$\psi_1(x)=\psi_0([\theta_0(t_1)][p+qx])$).  Using this new
$\psi_1(x)$, the same system of equations holds.

Figure~\ref{fig:intervene} compares these strategies.  As anticipated,
the final sizes resulting from cases 2 and 3 are identical, regardless
of the time of intervention.  However, we see that the dynamics are
significantly different.  
\begin{figure}
\includegraphics{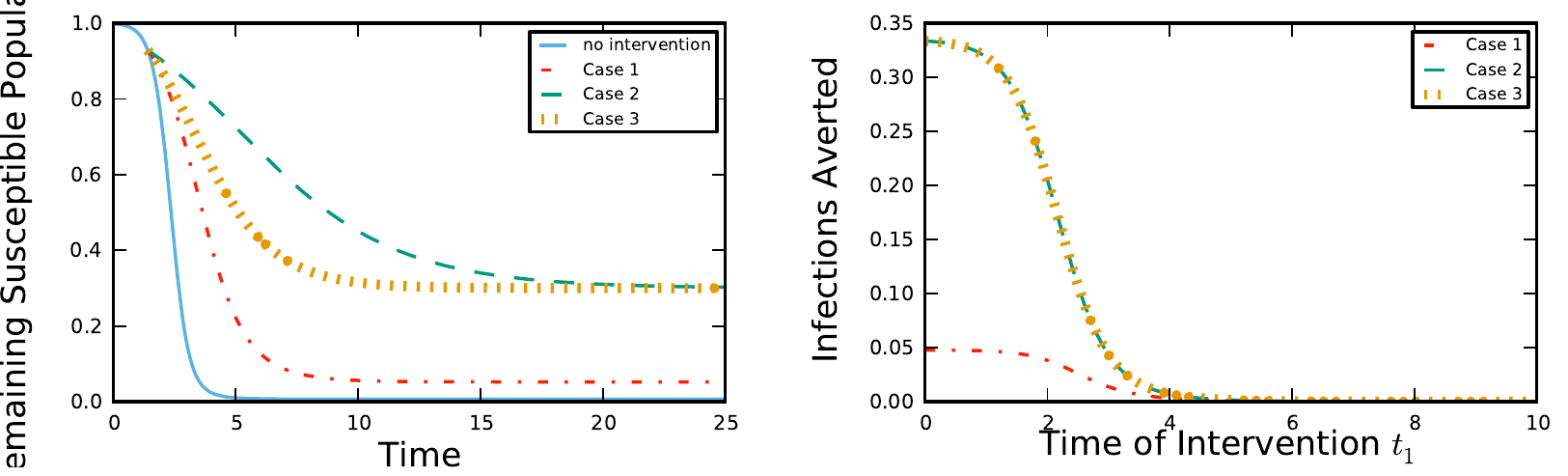}
\caption{\small The impact of interventions.  Epidemics begin at $t=0$ with
  $0.001$ of the population infected.  Left: epidemic curve without
  interventions, and with each intervention introduced at time
  $t_1=1.5$.  Right: horizontal axis is $t_1$, showing final
  effectiveness if interventions introduced at different times.}
\label{fig:intervene}
\end{figure}

\subsection*{Bifurcation analysis}

We try to gain a better understanding of the epidemic transition and
what happens to the final size as the initial proportion infected is
increased.  Consider the final size relation found from
\[
\theta = \frac{\beta\left[\phi_S(0)\frac{\psi'(\theta)}{\psi'(1)} + \phi_R(0)\right] +
\gamma}{\beta+\gamma}
\] 
If $\phi_S(0) + \phi_R(0)=1$, then we find that $\theta=1$ is a
solution to these equations.  Physically this states that if there is
no infection initially [$\phi_I(0)=0$] there will be no infection
later.  To study what happens when $\phi_I(0)>0$ (but possibly
arbitrarily small), we begin by first analyzing the structure of the
dynamical equation for $\theta$ under the assumption that
$\phi_S(0)=1$ and  $\phi_R(0)=0$, taking $\theta<1$.  These
assumptions contradict our initial conditions, but understanding this
system first will lead to an easier understanding of the full system
with $\phi_I(0)>0$.

\begin{figure}
\includegraphics{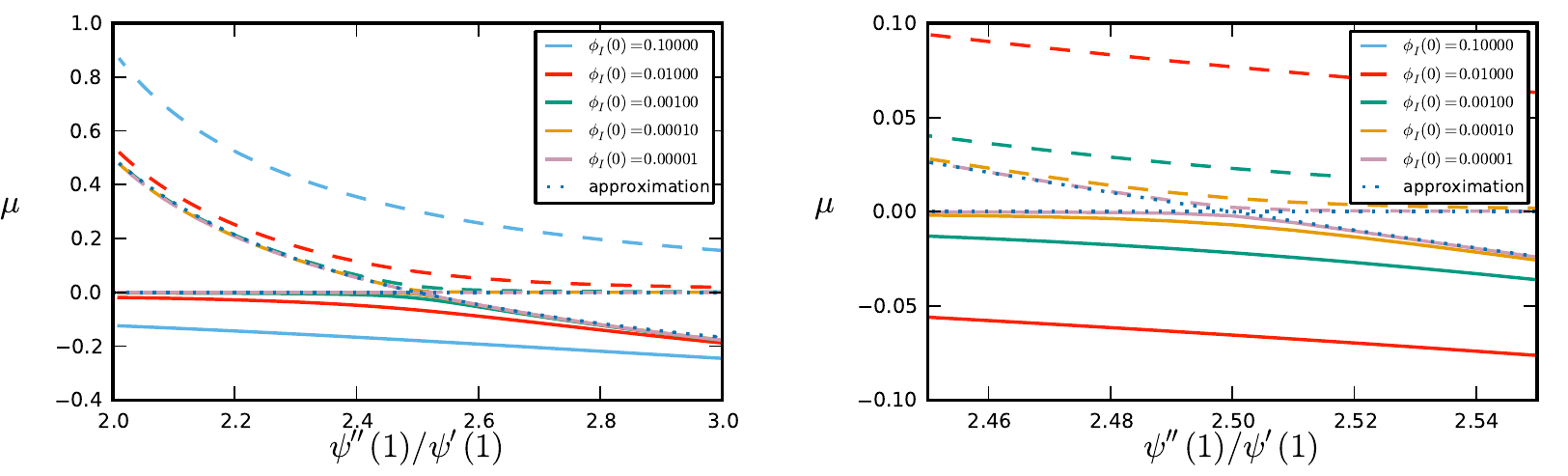}
\caption{\small Bifurcation diagram with $\psi_R(0)=0$, \
  $\phi_I(0)=1-\phi_S(0)$, and $\phi_S(0)$ as given.  Disease parameters are
  $\beta=1$ and $\gamma=1.5$.  In each all members of the population
  have degree either $3$ or $4$, with the proportions chosen so that
  $\psi''(1)/\psi'(1)$ takes the values on the horizontal axis.
  Approximate curves come from
  equation~\eqref{eqn:muapprox}.  Only the equilibria with $\mu<0$ are
physically meaningful.}
\end{figure}

If $\phi_S(0)=1$ and $\phi_R(0)=0$, the equation for $\theta$ becomes
\[
\dot{\theta} = -\beta \theta + \beta 
\frac{\psi'(\theta)}{\psi'(1)} + \gamma(1-\theta) 
\]
which has $\dot{\theta}=0$ whenever
\[
\theta = \frac{\beta \frac{\psi'(\theta)}{\psi'(1)} + \gamma}{\beta+\gamma}
\]
Clearly $\theta=1$ is an equilibrium.  Close to $\theta=1$, we write
$\theta = 1+\mu$, so $\psi'(\theta)/\psi'(1) = 1 + \mu
\psi''(1)/\psi'(1) + \mu^2 \psi'''(1)/2\psi'(1) + \order(\mu^3)$.
Substituting this into the equation for the equilibrium we have
\[
1+\mu = 1 + \mu\beta \frac{ \psi''(1) + \mu \psi'''(1)/2 + \order(\mu^2)}{\psi'(1)(\beta+\gamma)}
\]
which yields 
\begin{equation}
\mu= 0 \qquad \text{or} \qquad \mu \approx \frac{2}{\psi'''(1)} \left[
  \frac{\psi'(1)(\beta+\gamma)}{\beta} - \psi''(1) \right]
\label{eqn:muapprox}
\end{equation}
So there is a bifurcation as the bracketed term passes through zero,
when $(\beta+\gamma)/\beta = \psi''(1)/\psi'(1)$.  This is the
well-known epidemic threshold~\cite{newman:spread}.  The bifurcation
is transcritical and corresponds to $\Ro$ increasing through $1$.  If
$\psi''(1)/\psi'(1)< (\beta+\gamma)/\beta$ (that is $\Ro<1$) then
$\mu$ is positive and the corresponding equilibrium has $\theta>1$ and
is unstable, while the equilibrium at $\theta=1$ is stable.  In our
case, we will not observe $\theta>1$ because $\theta$ is a
probability.  If however $\psi''(1)/\psi'(1)> (\beta+\gamma)/\beta$
(that is $\Ro>1$), then the corresponding equilibrium has $\theta<1$ and
is stable while the equilibrium at $\theta=1$ is unstable.

We are now able to consider the effect of realistic initial
conditions.  We keep $\phi_R(0)=0$, but take $\phi_I(0)$ to be a small
positive number with $\theta(0)=1$ and $\phi_S(0) = 1-\phi_I(0)$.  The
bifurcation diagram changes slightly.  Compared to the equations
assuming $\phi_S(0)=1$, this has the effect of decreasing
$\dot{\theta}$ slightly, so the equilibrium values shift.

Below the bifurcation, the equilibrium $\theta=1$ is slightly reduced
to a $\theta_0<1$, but remains stable.  The solution with initial
condition $\theta=1$ converges to this equilibrium.  Above the
bifurcation, the $\theta=1$ equilibrium is slightly increased to a
$\theta_0>1$ and is unstable.  The other equilibrium with smaller
$\theta$ is stable and its location is decreased slightly.  Thus the
solution with initial condition $\theta=1$ decreases and converges to
the stable solution.

This bifurcation diagram helps explain some of the apparent
discrepancies of the earlier models.  When $\Ro<1$ and $\theta<1$, the
earlier models suggested that $\dot{\theta}>0$, that is, the
probability a neighbor has transmitted reduces in time.  This results
from the fact that the model assumed a negligible initial proportion
infected, and so it could not capture the fact that the stable
equilibrium at $\theta=1$ is reduced slightly by the initial
condition.  In these equations, the system converges to an equilibrium
that is in the wrong place.  Similarly, when $\Ro>1$, but the initial
proportion infected is not negligible, the system again converges to
an equilibrium in the wrong position.  This is what happened in
the~\cite{lindquist} paper which started with a very small initial
condition, and observed a very small discrepancy in the final size.



\subsection*{Conditions leading to failure of model}
There are some assumptions implicit in our derivation which deserve
further attention.  The model fails if $\theta(0)$, $\phi_S(0)$,
$\phi_I(0)$, or $\phi_R(0)$ depend on degree of $u$.  So
if for example, we select high degree individuals and then infect
their partners (leaving the high degree individuals uninfected), the
model will not account for the fact that higher degree individuals are
more likely to have infected partners at $t=0$.  The approach will
fail.

It does not fail if the initial individuals infected have higher (or
lower) degree.  This simply affects the initial conditions.  Indeed, we
expect that if the infection is initially spreading stochastically in the
population, and we set $t=0$ to be when enough cases are infected to have
deterministic behavior, we will see that at $t=0$ a disproportionate
number of higher degree individuals have been infected.  This does not
present a challenge.

\subsection*{Discussion}
We have shown that recent techniques used to derive epidemic dynamics
in networks may be adapted to situations in which the initial
condition is not small.  The resulting equations are relatively simple
to solve numerically.  We have shown how these equations can be used
to derive a final size relation.  Our results correct an apparent
discrepancy seen in earlier work comparing equations
of~\cite{volz:cts_time} with simulations of~\cite{lindquist}.  

One of the most obvious applications of these results is to the
understanding of the impact of an intervention which begins after a
disease has established itself.  This has been a weakness of network
models for some time: the earliest models could only calculate static
quantities such as the final size of epidemics assuming no
intervention, while more recent approaches that calculate the
dynamics~\cite{volz:cts_time,miller:volz,miller:ebcm_overview} have
been restricted to the assumption of asymptotically small initial
conditions, again with no change in the population.  Because we now
have a model which can account for large initial conditions, we can
use this to restart our calculations when an intervention is to be
implemented, or we can use the final size relation to quickly compare
intervention effectiveness.

We have analyzed the bifurcation structure of the final size
relation, and used this to explain an apparent discrepancy in earlier
work if $\Ro<1$.  The previous models that assumed small initial
condition also implicitly assume that $\Ro>1$.  This resulted in a
disturbing prediction for $\Ro<1$ that transmissions could be reversed
as time progresses, and infected individuals are uninfected.  Once we
correctly account for the initial condition this apparent discrepancy disappears.

If variance is large enough, then there may be a small number of very
high degree individuals who have a macroscopic effect on the dynamics.
If we increase population size to ``drown out'' their signal, we
expect to have a small number of much higher degree individuals who
again have a macroscopic effect on the dynamics.  Deterministic
predictions will not be accurate: for example, how long the highest
degree individual remains infected will influence the final size.  The
work of~\cite{decreusefond:volz_limit} rigorously studied the
equations using small initial conditions, and showed that if all
moments up to the fifth moment were finite, then these equations are
accurate in the limit of a large network.  Whether all these moments
are necessary is unclear (the equations are well-behaved so long as
the second moment is finite).  Regardless, when the equations do not
work, it is due to high degree individuals, so if the high degree
individuals are removed from the population, the equations will work.

After the epidemic has run for a short period of time, all of these
high degree individuals have been infected and recovered.  The
remaining population will have significantly reduced moments.  At this
point, the stochastic effects are ``frozen in'': the dynamics are now
deterministic.  We can use these conditions to initialize our new system
of equations.

\section*{Conclusions}

Recent advances in our understanding of infectious disease spread in
networks have allowed us to accurately predict SIR disease spread in a
range of networks, under the assumption of a negligibly small initial
condition.  However, in many contexts, such as might occur when an
intervention is applied, the small initial condition assumption
is false and the models give inaccurate predictions.  The method we
have introduced allows us to modify the previous equations and
eliminate the assumption that the initial condition is small.

Our modeling approach accurately predicts the size and dynamics of
simulated epidemics with arbitrary sized initial conditions.  The
approach allows us to compare interventions introduced during the
epidemic, which is not possible with previous network-based
approaches.

Our system of equations~\eqref{eqn:sys1} and~\eqref{eqn:sys2} are
mathematically simple and can be solved numerically with standard
tools.  Changes in the population's degree distribution do not alter
the structure of the equations, and in particular, the population can
have arbitarily large maximum degree without requiring any increase in
the number of equations.

\section*{Competing Interests}
The author declares that he has no competing interests
\section*{Authors Contribution}
This work is entirely the work of JCM.

\section*{Acknowledgements}
This work was supported by 1) the RAPIDD program of the Science and
Technology Directorate, Department of Homeland Security and the
Fogarty International Center, National Institutes of Health and 2) the
Center for Communicable Disease Dynamics, Department of Epidemiology,
Harvard School of Public Health under Award Number U54GM088558 from
the National Institute Of General Medical Sciences.  The content is solely the
responsibility of the authors and does not necessarily represent the
official views of the National Institute Of General Medical Sciences or
the National Institutes of Health.  The funding bodies had no role in
the design of this research, the writing of the manuscript, or the
decision to submit for publication.

{\ifthenelse{\boolean{publ}}{\footnotesize}{\small}
 \bibliographystyle{plain}  
  \bibliography{NetworkEpidemics} }     


\ifthenelse{\boolean{publ}}{\end{multicols}}{}

\end{document}